\documentclass[conference]{IEEEtran}
\IEEEoverridecommandlockouts
\usepackage{cite}
\usepackage{amsmath,amssymb,amsfonts}
\usepackage{algorithmic}
\usepackage{graphicx}
\usepackage{textcomp}

\usepackage{color, colortbl}
\definecolor{Gray}{gray}{0.9}
\usepackage{float}

\usepackage{graphicx}
\usepackage{xcolor}
\usepackage{multirow}
\renewcommand{\theenumi}{\roman{enumi}}%

\makeatletter
\def\ps@IEEEtitlepagestyle{%
    \def\@oddfoot{\mycopyrightnotice}%
    \def\@evenfoot{}%
}
\def\mycopyrightnotice{%
}
\let\old@ps@IEEEtitlepagestyle\ps@IEEEtitlepagestyle
\def\confheader#1{%
    \def\ps@IEEEtitlepagestyle{%
        \old@ps@IEEEtitlepagestyle%
        \def\@oddhead{\strut\hfill#1\hfill\strut}%
        \def\@evenhead{\strut\hfill#1\hfill\strut}%
    }%
    \ps@headings%
}
\makeatother

\def\BibTeX{{\rm B\kern-.05em{\sc i\kern-.025em b}\kern-.08em
    T\kern-.1667em\lower.7ex\hbox{E}\kern-.125emX}}
\begin{document}

\title{DNS attack mitigation Using OpenStack Isolation\thanks{}}

\author{\IEEEauthorblockN{Hassnain ul hassan
}
\IEEEauthorblockA{International Islamic University\\
% Kuala Lumpur, Malaysia\\
hassnainiqbal@gmail.com}
\and
\IEEEauthorblockN{Rizal Mohd Nor}
\IEEEauthorblockA{International Islamic University\\ 
% Kuala Lumpur, Malaysia\\
rizalmohdnor@iium.edu.my}
\and
\IEEEauthorblockN{Md Amiruzzaman}
\IEEEauthorblockA{West Chester University\\ 
% West Chester, PA 19383, USA \\
mamiruzzaman@wcupa.edu}
\and
\IEEEauthorblockN{Sharyar Wani}
\IEEEauthorblockA{International Islamic University\\ 
% Kuala Lumpur, Malaysia\\
sharyarwani@iium.edu.my}
\and
\IEEEauthorblockN{Md. Rajibul Islam}
\IEEEauthorblockA{
University of Asia Pacific\\
% Dhaka 1205, Bangladesh \\
md.rajibul.islam@uap-bd.edu}

% \and
% \and
}

\maketitle

\begin{abstract}
The Domain Name System (DNS) is essential for the Internet, giving a mechanism to resolve hostnames into Internet Protocol (IP) addresses. DNS is known as the world’s largest distributed database that manages hostnames and Internet Protocol. By having the DNS, only simple names that can be easily memorized will be used and then the domain name system will map it into the numeric Internet Protocol addresses that are used by computers to communicate. This research aims to propose a model for the development of a private cloud infrastructure to host DNS. The cloud infrastructure will be created using the OpenStack software platform where each server will be hosted separately in a different virtual machine. Virtual network architecture will be created using the Software Defined Networking (SDN) approach and it will be secured using Firewall as a Service (FWaaS). By hosting DNS in private cloud infrastructure, the DNS servers will be out of reach by attackers which will prevent DNS attacks. Besides, available research had proven that the cloud is the best choice for DNS. A prototype had been implemented and evaluated for its efficiencies. The findings from the evaluation carried out shown a positive result.

% Geo-coded multimedia data including spatial videos, speech, and geo-narratives are frequently used by domain users and practitioners in urban planning, social study, and public health applications. While the data can be captured by mobile devices, it is not easy to be visually explored at the same time. In this paper, we develop a mobile system which collects geo-coded multimedia datasets and directly process and analyze them with interactive visual exploration. This mobile system also applies AI-based semantic image segmentation and audio transcription to process the raw data, and them utilizes the results as cues for effective data analysis. The utilization of the software can span across disciplines as diverse as urban study and planning, geography, health, sociology, and education. 

\end{abstract}


\section{Introduction}
\label{sec:intro}
The threats that surround the Domain Name System (DNS) due to the lack of authenticity and integrity checking of the data held within the DNS components are often subjected to Denial-Of-Service (DOS) attacks. The DOS attacks intended to disrupt access to the resources whose domain names are handled by the attacked DNS components \cite{r1}. Hundreds and millions of users need DNS to translate domain names to IP addresses and back \cite{r2}. Therefore, they can access the Internet resources by user-friendly domain names rather than IP addresses, but any breach in DNS security will result in loss of trust to the whole internet. Therefore, the security of DNS is paramount. DNS is designed to be a public database \cite{r3}. Thus, wrong information provided by a DNS services may cause dire and possibly dangerous exposures.

Most of the weaknesses within the DNS fall into one of the following categories: cache poisoning (Davidowicz, 1999), client flooding, dynamic update vulnerability, information leakage, and compromise of the DNS several authoritative databases. One of the most discussed concepts in DNS related security is “Domain Name System Security” (DNSSEC). It is a DNS security extension to overcome DNS spoof attacks \cite{r4}. DNSSEC has an implementation complexity owing to key management issues. This is because zone signing requires it to be signed by a key and the key used must be signed as well. It is seeded at the top-level domain. Hence, DNSSEC does not provide confidentiality and this technique causes overheads in response size. A key mismatch may take down the whole DNS \cite{r4}. 

Anycast is a network routing and addressing method which is easily scalable \cite{r5}. One of the limitations for Anycast is the unawareness of the network conditions and server load. Many server-load AnyCast mechanisms have been proposed. A routing controller is deployed initially, then this route controller first acquires knowledge of network conditions and server loads from “Provider edge” (Pe) and “Content Delivery Network” (CDN) servers. This kind of approach is free of drawbacks, but it only works with a single autonomous system. With this CDN’s may get connectivity from numerous ISPs and their platforms are split among many autonomous systems \cite{r5}. 

Like other variations for DNS, the Namecoin was designed to stand against the legal attacks \cite{r6}. Implementing solution like their regulation is significantly more difficult, as the Blockchains does not have a legal entity associated with it. Namecoin can improve user privacy if the full blockchain is replicated at the user end. Replicating the complete blockchain at each user may be impractical for some devices. Another drawback is that Namecoin does not protect the zone information from monitoring \cite{r6}. 
The objective of this paper is two folds. First, propose a model that consists of secure scalable cloud infrastructure and workflow to be used for DNS hosting on a private cloud using OpenStack. Second, prevent attackers from approaching the main components of DNS using the proposed model mentioned earlier.

\section{Related Work}
\label{sec:relatedwork}

The standard DNS does not provide any protection from malicious sites. The techniques to crack DNS keep getting better and DNS is not able to resist to these attacks \cite{r7}.

\subsection{Most Known Attacks}
“DNS Spoofing” is the action of answering a DNS request that was intended for another server \cite{r8}. A form of computer security hacking in which corrupt Domain Name System data is introduced into the DNS resolver's cache, causing the name server to return an incorrect result record \cite{r13}, e.g. an IP address. 

Most of the TCP/IP suite protocols does not provide solutions for authenticating the source or destination of a message and in result vulnerable to spoofing attacks, where extra safety measures are not taken by applications to verify the authentication of the sending or receiving host. IP spoofing and ARP spoofing may be used to lead to man-in-the-middle attacks against hosts on a computer network. Spoofing attacks that take advantage of TCP/IP suite protocols may be mitigated with the use of firewalls capable of packet inspection or by taking measures to verify the identity of the sender or recipient of a message \cite{r13}. 

“DNS ID Hijacking” \cite{r9} is when DNS uses the ID number to identify queries and answers, and the hacker finds the ID the client is waiting for. With DNS spoofing, the hacker will try to impersonate the DNS reply so that their questing client is misdirected, but without touching the DNS cache of the impersonated DNS.
“DNS cache Poisoning” happens when a DNS cache can become poisoned when the attackers take over DNS and then insert false information and then direct the user to a different address that functions as a phishing website \cite{r14}.

\subsection{Current Methods}
Maksutov, Cherepanov, and Alekseev \cite{r8} wrote a paper about attacks done by taking advantage of DNS exploits. Besides that, they described a simple tool to protect against replacing DNS responses. However, the objectives of this study is on the detection and prevention of DNS spoofing attacks using a tool called DNSwitch. The authors have clearly stated that using this proposed technology will not prevent all DNS-spoofing attacks but helps to protect it without putting too much load on the computer. DNSwitch during the DNS query creates two copies of their request, one of which will be sent to the local DNS server, and the second one is sent through a secure channel to the trusted server and later on it will be used to authenticate the response from the local server. 

Benshoof et. al. \cite{r10} presented a paper in using the Bitcoin blockchain to create a distributed DNS that replaces the current Top-Level Domain Name System. The research aims to overcome security issues and scalability. Besides that, it tries to eliminate the need for certificate authorities through a decentralized authenticated record of domain name ownership. The blockchain does not solve all security issues relevant to DNS authentication and security. D3NS uses Distributed Hash Tables (DHT) which is not a good choice for searching due to the consequences of the hashing algorithm. Furthermore, the system complicity and reliance on a lot of new untested technologies introduce a threat where it will be very difficult to ensure efficiency. 

Buijsman, Mekking, and Ham \cite{r11} wrote a research paper on TSIG, which is used mostly during DNS updates and zone transfer and uses private key encryption to secure the last mile of DNS messages between the stub resolver and the recursive resolver. To generate a receiver host's signature to examine the message's reliability based on a shared password, the TSIG uses a one-way handshake. TSIG is a protocol that is specified in "RFC2854" and it provide DNS messages with authentication. However, it is a limited solution to securing DNS. It depends on the manual distribution of secret key, which can be unsafe and challenging to implement as large percentage of clients serve the recursive resolver, and all users need to get the secret key installed manually. Another essential drawback to TSIG is the lack of confidentiality. It also deals with the private key's automated distribution instead of the manual method, but this protocol expansion does not resolve the confidentiality problem.

\section{PROPOSED METHOD}
\label{sec:proposedmethod}
As aforementioned in the previous section, there are some disadvantages to some existing solutions. DNSwitch does not prevent all DNS-spoofing attacks, but it just helps to protect it without putting significant load on the computer \cite{r8}. Also, D3NS does not deal with all the security issues relevant to DNS authentication and security \cite{r8}. Moreover, DHT is not a good choice for searching due to the consequences of the hashing algorithm. Additionally, the system complicated, and untested nature makes it difficult to ensure efficiency \cite{r10}. 

The DNSCrypt acts as a stand-alone proxy and the mechanism that is based on DNSCURVE, which has yet to be standardized \cite{r12}. The TSIG is a weak solution for securing DNS. It depends on the manual distribution of the secret key, which is unsafe and challenging to implement, realizing that a large percentage of clients serve the recursive resolver, and all users need to get the secret key installed manually. Another essential drawback to TSIG is the lack of confidentiality.  
The GSS-TSIG also deals with the private key's automated distribution instead of the manual method, but this protocol expansion does not resolve the confidentiality problem \cite{r11}. 

In a study, Aishwarya \cite{r12_1} justified that the cloud is the best practice for distributive data security, and it is a security mechanism from external sources. Besides, the author described the advantages that make the cloud the best choice for hosting the DNS servers \cite{r12}. Using the cloud to host the DNS servers will overcome those issues by simply using the existing DNS hosted in a private cloud and will not need to bring up new untested and untrusted technologies that are difficult to understand. Also, no configuration or installation of any additional system needed by the clients \cite{r12}. 

In this study, we proposed the use of a DNS hosting model in a secure private cloud environment to overcome the security issues that DNS is facing for the past few years. As per the definition outlined by the "National Institute of Standards and Technology” (NIST), Cloud computing is a base design that enables a convenient, ubiquitous, and on-demand network access to shared computing resources that are configurable and can be rapidly provisioned and released with minimal management effort or extensive interaction with a service provider \cite{r12}. 

There are several reasons to choose the cloud from a security perspective. These reasons are extensively mentioned by Lafta and M.and Mihaylov \cite{r12} paper which includes that the cloud is good in managing distributive data. The cloud by itself is distributed in nature and DNS is also a distributive data service and hence makes it is easy to create and manage DNS servers on the cloud. The main concern of DNS security is securing the system from external sources. Though data availability in DNS Servers is important, it is not as sensitive as other types of data such as financial data, medical data, etc. 

The only threat related to the security of the cloud is from the cloud service providers and the government under whose jurisdiction the server is placed. No external entity without the permission of the cloud service provider and the owner can access the cloud \cite{r12}. This is in our justification, makes cloud the best place to host DNS. However, the proposed model consists of the following parts which is Isolating DNS components in a private cloud and Mapping domain names to IP addresses and back.

\subsection{Development}
The initial setup starts with the physical BareMetal servers that was prepared with the Ubuntu operating system and pre-installations to make sure the operating system is stable on the server. Subsequently, all the required packages and sources were installed to assure OS stability. The next step was setting up network connectivity configuration, between those servers by creating different VLANs, using the approach of “Software Defined Networking” (SDN). This step involves manual tinkering to the hardware resources via the terminal. This is the core part of our DNS Isolation. After this stage, the model was divided into two phases: first phase is building a private cloud infrastructure by setting up Kernel Base Virtual Machine (KVM) and the second phase is to host the DNS servers inside the created VMs and make the configuration needed for DNS to work. 

There is a need to create a VM for Security Monitoring, and this model uses PfSense. The PfSence is an opensource firewall and most stable software distribution, based on FreeBSD. From here we can create different users to manage the network security for our BareMetal level. User-level users are created to control the VMs. For the second phase, OpenStack-Ansible was installed in one of the VMs to manage KVM. This is required to help manage other VMs. OpenStack-Ansible makes it very easy to manage all the hardware Resources and VMs Through OpenStack Dashboard.

Once the cloud infrastructure is ready the DNS hosting environment was created by configuring the needed security and network followed by creating four VMs for DNS hosting. Each VM to function different DNS servers. Then the DNS servers are torn apart and ran in different VMs. For the query of any domain search, DNS servers will talk through the VM level. All the traffic will be monitored through OpenStack. 
    
\subsection{DDoS within Cloud}
There are many types of DDoS attacks, but for the purpose of this research we will discuss two types of DDoS attack which are 1) “TCP-sync flood” which are DDoS within the cloud and 2) “Low-frequency DDoS attack” (LDDoS).

In “DDOS within cloud”, many studies have been done on the Intrusion Detection System, and the DoS attacks mechanism on a host occupying the cloud, while the attacker is situated externally from the cloud. Nevertheless, because of the cloud environment's multi-tenancy nature, malicious tenants can harm the other cloud's legitimate tenants, especially for cloud deployed as a MaaS model from all the tenants residing on the cloud. DDOS attacks will be explored internally and externally. Dos attack is operated on a VM running a web server in a TCP SYN flood attack. The web server has web server cookies unbaled to TCP SYN, which is usually asserted to secure the webservers from the flood attacks of TCP SYN. Moreover, the DoS's external network attacks can be managed by implementing the same approach; however, during experimentation, all attacks are performed by VMs within the cloud.

One type of DDOS attack is the “LDoS attack on cloud computing platform” which is a low-frequency DDoS attack. It is mainly used to determinate the traffic from attackers to non-attackers \cite{r15,r16,r17}. In this research the following essential observations are noted as follows.

\begin{itemize}
    \item The legitimate clients cover a larger geographical area compared to the attack client’s area. Therefore, the distribution of source IP is more diffused for regular traffic than the distribution of source IP for attack traffic

   \item The size of non-attack packets is according to the pattern as per the client's request. However, this behavior is not available in attack packets as attack clients usually generate equal size packets. Hence, the packet size of distribution for non-attack flow is comparatively more diffused than the packet size of distribution for attack flow.
\end{itemize}

“Sniffer and Feature Extraction”. Virtualization on the NC allows a bridged interface to be set up, i.e., the br0 interface. All local VM interfaces are therefore connected to the formed interface so external networks can be communicated quickly. Optimally, all packets arriving from the external network via the bridged interface are routed with the iptables firewall inside Linux's operating system. The OpenStack safety feature which permits VMs to be allocated to a security group includes the rules to be applied to the security group as firewall rules in the local NC firewall iptables.

The cloud is being deployed with OpenStack "version 16" software, which operates on three physical servers and has a separate Storage Area Network (SAN), which results in a processing capacity of 20 VCPUs" and storage space of "7.3 TeraByte".  Each server has a basic version of Ubuntu 18 installed, and the cloud components are mounted on the top of the operating system. Deployment of every OpenStack component is performed on each physical server that provides these components with dedicated server facilities to execute the functions. The SAN is used as a block storage method for the Storage Controller (SC). The cloud uses the KVM hypervisor to virtualize the processing, storage services, networking, and deliver IaaS. 

OpenStack cloud must be designed in a way that allows components to interact with one another. When launched, the VMs should connect and the external network with absolute security. The Edge networking mode used in this configuration avoids the need to put a single Linux server in the data path for all VMs operating in a single cluster, eliminating the need to configure the underlying network to enable packets tagged to VLAN in the cloud.

A list of private and public IP addresses must also be assigned to OpenStack. OpenStack incorporates a function known as cloud security groups. Security groups serve as virtual firewalls that help ingress packet filtering for VMs on which they are added. Security groups are enforced from the CLC to VMs and are used as iptables guidelines for NCs. Two security groups are set up for VMs on this testbed: one for a user and attack VMs, and the other for VMs on the webserver.

\section{EXPERIMENT}

“Experiment for DDoS attack” is based on the main objective set for this paper, the proposed DNS hosting model was evaluated as hosting DNS in a secure private cloud and to prevent attackers form reaching to the main components of it. The aim of this test was to determine the effectiveness of this model, the security and the speed of the query. To test the speed and performance of the proposed model, two tools was used beside using the normal “nslookup” for DNS query:
\begin{itemize}
    \item Hping3 is used during testing to send infinite packets to the DNS server through DNS port to overload it with queries.
    \item Namebench is used to test the proposed model and comparing it to OpenDNS and google DNS.
\end{itemize}

\begin{figure}[h]
	\centering
	\includegraphics[width=0.4\textwidth]{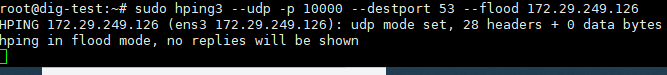}
	\vspace*{-3mm}
	\caption{LDoS attack pattern representation}
	\label{fig:01}
\end{figure}

“Load and Speed Evaluation” data are conducted using “Hping3” and “Namebench”. As shown in Fig. \ref{fig:01}, Hping3 is executed at first and kept running to keep flooding the DNS server with queries instead of keeping the server idle. Continuous flooding allows us to collect and evaluate more effectively.

%%%% Fig 1

The input flood of DNS messages is set to the following command switches, “udp” for “udp” Datagram protocols, port 53 for “destport” and DNS server IP address to simulate DNS flood data. Concurrently, we ran and monitored our proposed DNS with simulated flooding using Hping3 and used “Namebench” to capture and monitor our experiment for speed and performance of the proposed model.	

In our “Testbed for DDoS attack within cloud” various functions are delegated to VMs. Because the research focuses on VM communication from VM and attacks, it is essential to simulate an active cloud environment where VM communication is necessary. Three different VMs are set up in the cloud but they randomly request webpages from the VM webserver. The VM that runs an instance of the web server running "Apache2.2.15" is hosting a website of 6-pages. The instances are set up as VM attacks, where TCP SYN flood attacks are tasked to perform attacks based on scenarios and time allocated.
The “Siege” is a web server load test and benchmarking tool, that was used in our experiment. The Siege parameters were set so that Siege would request each of the six web pages from the webserver at random times within a specified time for user 'VM 1' to 'VM 10'. For instance, Siege will request a website from the web server for user "VM 4" at any moment between 20 seconds and later request another webpage within next 20 seconds.

%%% Fig 2

\begin{figure}[h]
	\centering
	\includegraphics[width=0.4\textwidth]{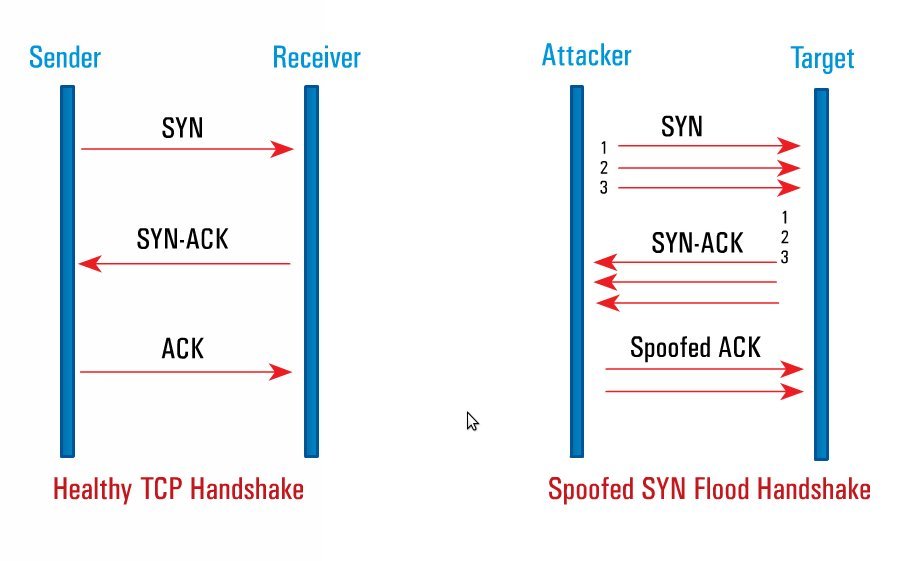}
	\vspace*{-3mm}
	\caption{LDoS attack pattern representation}
	\label{fig:02}
\end{figure}

“TCP-Syn Flood” uses the TCP protocol's 3-way handshake link establishment mechanism to execute the attack. In a typical case, the client sends the SYN packet to the server as a request to connect to the server, after which the server accepts the reception of the request and sends the SYN and ACK bits to the client. The server will have allocated resources for this client at this stage. After sending the ACK back to the computer, the recipient completes the handshake. 

In a TCP SYN flood, the client sends several SYN packets to the server at a very high rate. With either its IP addresses or spoofed addresses. Fig. \ref{fig:02}. illustrates the difference in interaction between the sender and receiver for TCP SYN floods. Once the server receives these SYN packets, the server allocates resources and sends a packet with SYN and ACK bits set to either the recipient or spoofed addresses. The client or spoofed addresses does not react with an ACK packet to the server. The hping3 tool is used to execute TCP SYN flood attacks, enabling the server to send customizable SYN packets. For a total duration of 4 hours, four attack scenarios are carried out, and the details are outlined in “Table \ref{tbl:1}”.

“Results from DDoS attack” in a four (4)-hour experimental study involving user VMs that sent requests that are legitimate and attack VMs that sent malicious requests to the VM webserver, the sniffer’s outcome, and the extraction process culminated in a dataset of "5026" instances "x 9" features. The distinction between the two types of legitimate and malicious traffic from VMs was that "10 VMs" sent legitimate traffic. Although only "2VMs" sent malicious traffic, the sniffer and the extraction process's function was focused on the incoming IP address in the traffic flow to catch and extract features from it for a designated time interval of 5 seconds. The primary dataset is segregated into a ratio of "4:1" which represents the test dataset and training groups.

After splitting, the dataset of training consists of "4021" instances from which legitimate class was numbered to "3513" instances, and the remaining "508" instances belong to the malicious class. The test data set includes "1005" instances consisting of total "878" \& "127" instances belonging to the legitimate and malicious classes, respectively. The extract of the original data set is seen in Table \ref{tbl:2}. It includes "10" instances "x 9" features and an extra column that identifies instances as either legitimate or malicious.

\section{DISCUSSION}

The objective of any DoS attack is to exhaust resources of the system so it should be unable to offer service to the clients \cite{r15}; the goal of the TCP SYN flood attack is the same. Looking at the output of a VM webserver, attacked by one and two VMs (simultaneously) can be known by calculating transactions from an arbitrary client requiring a service while the webserver is under attack and not in a normal situation. If there is no attack on the webserver but just legitimate requests from the VM user and the four situations where the webserver is under attack. 

However, experiments indicate no substantial improvements in these two system resources during traffic handling attacks. There has been a small rise in memory consumption when the TCP SYN flood attack was in progress for 5 minutes. According to the observation, there was no significant change contrary to normal memory use. The attacks do not significantly impact system loads. 

From the data table, the potential symptoms of TCP SYN flood attacks are higher incoming than normal outgoing network traffic. However, it differs from other cases that, in certain instances, which are simultaneously accessed by a large number of clients to a web server, any false positive identification may be created by an intrusion detection system.

\begin{figure}[h]
	\centering
	\includegraphics[width=0.4\textwidth]{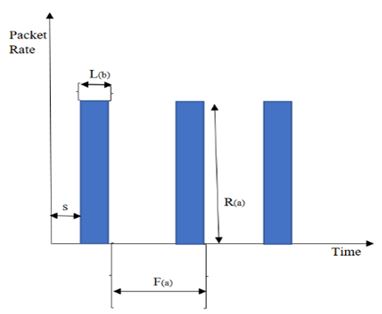}
	\vspace*{-3mm}
	\caption{LDoS attack pattern representation}
	\label{fig:03}
\end{figure}

“LDoS attack representation” in LDoS attacks, very low-frequency attack spikes, are sent by the attacker. The pattern of the LDoS attack can be represented using the following four parameters: F(a), L(b), R(a), and s. Here F(a) is the frequency of the LDoS attack, L(b) is the length of the attack burst, R(a) is the rate of the packets during the attack burst, and s is the starting time of the attack as shown in Fig. \ref{fig:03}.

“LDoS attack flow detection” refers to a statistical approach based on the hypothesis tests to detect the LDoS attack. Basic statistical concepts are followed to present the proposed approach. The following sequence of steps shall be taken to conclude the proposed test.

%%% Steps
\begin{enumerate}
    \item The hypotheses events are marked as follows: H(0) (null hypothesis) for legitimate traffic and H(1) (alternate hypothesis) for attack traffic. The hypothesis events are define as follows:
    \begin{equation}
        H(0): \sigma_N = \sigma_r
    \end{equation}
     \begin{equation}
        H(1): \sigma_N = \sigma_R
    \end{equation}
    
    where $\sigma_N$ and $\sigma_R$ are considered as standard deviations of legitimate ($X_N$) and attacks in (real-time) traffic as ($X_R$) respectively.
    \item For the test statistic (T-statistic) measuring with the help of the following formula: 
    \begin{equation}
        t_{obs} = \frac{\bar{X}_N -\bar{X}_R}{\sqrt{\frac{\sigma^2_N}{n}+\frac{\sigma^2_R}{m}}}
    \end{equation}
    where $n$ and $m$ both are number of samples in $X_N$ and $X_R$ respectively.
    
    \item The non-acceptance area of tabs can be given as, $t_{obs} \leq t_{\alpha}(v)$. where $t_{\alpha}(v)$ is T-distribution's value with an $\alpha$ level of importance and $v$ is the degree of freedom defined by the Satterwaite approximation (formula \ref{eq:approximation})
    \begin{equation}
        v = \frac{ ( \frac{S^2_N}{n}+\frac{S^2_R}{m} ) }{ \frac{S^4_N}{n^2(n-1)}+\frac{S^4_R}{m^2(m-1)} }\label{eq:approximation}
    \end{equation}
    
    \end{enumerate}

\begin{table}[h]
\centering
\caption{Attack Scenario}\label{tbl:1}
\begin{tabular}{|l|l|l|l|}
\hline
No & Attack VMs & Data in Packet         & Attack duration            \\ \hline
1  & 1          & 40   bytes/packet      & 6x5   min attack for 1hour \\ \hline
2  & 1          & 1500   bytes/packet    & 6x5   min attack for 1hour \\ \hline
3  & 2          & 40   bytes/packet/VM   & 6x5   min attack for 1hour \\ \hline
4  & 2          & 1500   bytes/packet/VM & 6x5 min attack for 1hour   \\ \hline
\end{tabular}
\end{table}

\begin{table*}[!htbp]
\centering
\caption{Results Captured}\label{tbl:2}
\begin{tabular}{|l|l|l|l|l|l|l|l|l|l|}
\hline
Bytes\_In & Avg\_Bytes\_in & Count  & AvgCount & All Count & BytesOut & AvgBytes\_out & SynBit & AckBit & Outcome    \\ \hline
1457      & 2005.8         & 5      & 6.6      & 33        & 437      & 542.4         & 1      & 5      & Legitimate \\ \hline
3991      & 2409           & 11     & 7.333    & 22        & 874      & 578.33        & 4      & 11     & Legitimate \\ \hline
2047      & 952856         & 6      & 21649    & 194849    & 373      & 4.75E+07      & 1      & 6      & Legitimate \\ \hline
1612      & 2154.5         & 3      & 5.25     & 21        & 332      & 443.5         & 1      & 3      & Legitimate \\ \hline
7923      & 4259.33        & 23     & 13       & 39        & 1747     & 1074          & 4      & 23     & Legitimate \\ \hline
7446460   & 1.72E+06       & 169244 & 39091.2  & 195459    & 1462032  & 3212179       & 169175 & 0      & Malicious  \\ \hline
2684      & 1430.25        & 61     & 18.75    & 75        & 0        & 218.5         & 61     & 0      & Malicious  \\ \hline
264       & 2402.6         & 6      & 8.2      & 41        & 0        & 558.4         & 6      & 0      & Malicious  \\ \hline
3453600   & 1921821        & 78497  & 43676.8  & 218384    & 8674440  & 3.72E+06      & 78430  & 0      & Malicious  \\ \hline
4540136   & 2330457        & 103185 & 52956.4  & 264782    & 9278560  & 3.49E+06      & 103184 & 0      & Malicious  \\ \hline
264       & 2402.6         & 6      & 8.2      & 41        & 0        & 558.4         & 6      & 0      & Malicious  \\ \hline
3453600   & 1921821        & 78497  & 43676.8  & 218384    & 8674440  & 3.72E+06      & 78430  & 0      & Malicious  \\ \hline
4540136   & 2330457        & 103185 & 52956.4  & 264782    & 9278560  & 3.94E+06      & 103184 & 0      & Malicious  \\ \hline
\end{tabular}
\end{table*}

\section{EXPERIMENTAL RESULTS}
For our experiments, we use data from the “Defence Advanced Research Projects Agency” (DARPA) dataset \cite{r16} and CAIDA \cite{r17} DDoS attack datasets. All packets included in the DARPA data set are legitimate, and all packets used in the “Center for Applied Internet Data Analysis” (CAIDA) data set are attack packets. These datasets have been used for several high-quality works in other previous research \cite{r18}. This is to eliminate doubts of using reliable datasets. The dataset from DARPA is used to simulate attack-free traffic. The data obtained from Tuesday of the first training week is is taken to simulate and select present normal network conditions for the experiments as it comprises only legitimate packets. The CAIDA \cite{r17} dataset reflects the attack traffic as it consists of approximately 300 seconds of attack traffic with the deleted identification of the DDoS attack registered on 4 September 2020. 

Legitimate packets are removed as far as possible from the dataset. The packet rates in the flooding DDoS traffic should be more than 10000 attack packets/second as per the literature's observation. 
The packets rate of the section of the CAIDA \cite{r17} dataset was selected for approximately 500 packets/second for the experiments. Therefore, this dataset is suitable to be considered as LDoS attack traffic, as stated in \cite{r18} To get the resulting dataset representing the real attack traffic, the selected portion of the DARPA and CAIDA datasets dataset is mixed as the LDoS attack includes both the attack packets and valid packets.

% Other results obtained are the probability distribution of the network flow using IP packet-size which was computed to track the LDoS attack traffic. Standard deviation measuring the sum of the participants' deviation from its mean is also collected. A high standard deviation value implies a more significant deviation.

\section{CONCLUSION}
The research work addressed a crucial problem in a cloud environment concerning VM to VM DoS attacks and another form of DDoS attack known as LDoS. In this type of network, the MaaS cloud environment, itself the virtual representation of a network and cloud providers, needs to provide its users with a certain level of security. It is not enough to only protect client VMs from DoS attacks depending on OpenStack security groups. Instead, additional attacks such as the TCP SYN flood attack must be considered. It was found that with the help of strict rules of the security group, which enables SYN cookies on the webserver of a VM, the TCP SYN flood attack scan causes performance degradation of a web server VM running Apache. Furthermore, it renders it up to 10,100 times slower than its normal capacity to react to requests.

% \clearpage
\bibliographystyle{IEEEtran}      % IEEE
\bibliography{bibfile}

\begin{thebibliography}{10}
\providecommand{\url}[1]{#1}
\csname url@samestyle\endcsname
\providecommand{\newblock}{\relax}
\providecommand{\bibinfo}[2]{#2}
\providecommand{\BIBentrySTDinterwordspacing}{\spaceskip=0pt\relax}
\providecommand{\BIBentryALTinterwordstretchfactor}{4}
\providecommand{\BIBentryALTinterwordspacing}{\spaceskip=\fontdimen2\font plus
\BIBentryALTinterwordstretchfactor\fontdimen3\font minus
  \fontdimen4\font\relax}
\providecommand{\BIBforeignlanguage}[2]{{%
\expandafter\ifx\csname l@#1\endcsname\relax
\typeout{** WARNING: IEEEtran.bst: No hyphenation pattern has been}%
\typeout{** loaded for the language `#1'. Using the pattern for}%
\typeout{** the default language instead.}%
\else
\language=\csname l@#1\endcsname
\fi
#2}}
\providecommand{\BIBdecl}{\relax}
\BIBdecl

\bibitem{r1}
D.~Kapil, ``{Dnssecurity},'' \url{https://dhavalkapil.com/blogs/DNS-Security/},
  2015.

\bibitem{r2}
P.~Mockapetris \emph{et~al.}, ``Domain names-implementation and
  specification,'' 1987.

\bibitem{r3}
D.~Davidowicz, ``Domain name system (dns) security,'' \emph{Yahoo Geocities},
  1999.

\bibitem{r4}
B.~Rajendran, P.~Shetty \emph{et~al.}, ``Domain name system (dns) security:
  Attacks identification and protection methods,'' in \emph{Proceedings of the
  International Conference on Security and Management (SAM)}.\hskip 1em plus
  0.5em minus 0.4em\relax The Steering Committee of The World Congress in
  Computer Science, Computer~…, 2018, pp. 27--33.

\bibitem{r5}
B.~Yan, B.~Fang, B.~Li, and Y.~Wang, ``Detection and defense of dns spoofing
  attack,'' \emph{Computer Engineering}, vol.~32, p.~21, 2006.

\bibitem{r6}
C.~Grothoff, M.~Wachs, M.~Ermert, and J.~Appelbaum, ``Toward secure name
  resolution on the internet,'' \emph{Computers \& Security}, vol.~77, pp.
  694--708, 2018.

\bibitem{r7}
A.~Liska and G.~Stowe, \emph{DNS Security: Defending the Domain Name
  System}.\hskip 1em plus 0.5em minus 0.4em\relax Syngress, 2016.

\bibitem{r8}
A.~A. Maksutov, I.~A. Cherepanov, and M.~S. Alekseev, ``Detection and
  prevention of dns spoofing attacks,'' in \emph{2017 Siberian Symposium on
  Data Science and Engineering (SSDSE)}.\hskip 1em plus 0.5em minus 0.4em\relax
  IEEE, 2017, pp. 84--87.

\bibitem{r13}
D.~Malisencu, ``Attacks analyze in the computer networks,'' 2019.

\bibitem{r9}
A.~A.~Z. Hudaib and E.~Hudaib, ``Dns advanced attacks and analysis,''
  \emph{International Journal of Computer Science and Security (IJCSS)},
  vol.~8, no.~2, p.~63, 2014.

\bibitem{r14}
B.~Krueger, ``Using a certificate public key to protect dkim public key
  spoofing,'' 2020.

\bibitem{r10}
B.~Benshoof, A.~Rosen, A.~G. Bourgeois, and R.~W. Harrison, ``Distributed
  decentralized domain name service,'' in \emph{2016 IEEE International
  Parallel and Distributed Processing Symposium Workshops (IPDPSW)}.\hskip 1em
  plus 0.5em minus 0.4em\relax IEEE, 2016, pp. 1279--1287.

\bibitem{r11}
M.~Buijsman, M.~Mekking, and J.~van~der Ham, ``Securing the last mile of dns
  with cga-tsig,'' \emph{Research Project}, vol.~2, 2014.

\bibitem{r12}
M.~Lafta and G.~Mihaylov, ``Securing dnssec last mile with dtls,'' 2015.

\bibitem{r12_1}
C.~Aishwarya, M.~Sannidhan, and B.~Rajendran, ``Dns security: Need and role in
  the context of cloud computing,'' in \emph{2014 3rd International Conference
  on Eco-friendly Computing and Communication Systems}.\hskip 1em plus 0.5em
  minus 0.4em\relax IEEE, 2014, pp. 229--232.

\bibitem{r15}
J.~Idziorek, M.~F. Tannian, and D.~Jacobson, ``The insecurity of cloud utility
  models,'' \emph{IT Professional}, vol.~15, no.~2, pp. 22--27, 2012.

\bibitem{r16}
K.~Bhushan and B.~Gupta, ``Hypothesis test for low-rate ddos attack detection
  in cloud computing environment,'' \emph{Procedia computer science}, vol. 132,
  pp. 947--955, 2018.

\bibitem{r17}
Y.~Xiang, K.~Li, and W.~Zhou, ``Low-rate ddos attacks detection and traceback
  by using new information metrics,'' \emph{IEEE transactions on information
  forensics and security}, vol.~6, no.~2, pp. 426--437, 2011.

\bibitem{r18}
C.~Zhang, Z.~Cai, W.~Chen, X.~Luo, and J.~Yin, ``Flow level detection and
  filtering of low-rate ddos,'' \emph{Computer Networks}, vol.~56, no.~15, pp.
  3417--3431, 2012.

\end{thebibliography}
\end{document}